\documentclass[journal]{IEEEtran}
\makeatletter
\let\@Lbot\relax
\let\@Rbot\relax
\makeatother
\usepackage[pdftex]{graphicx,xcolor}
\usepackage[fleqn]{amsmath}
\usepackage{booktabs}
\usepackage{newtxtext}
\usepackage{float}
\usepackage{authblk}
\usepackage[varg]{newtxmath}
\usepackage{algorithm}
\usepackage{algpseudocode}

\title{Turning Threat into Opportunity: DRL-Powered Anti-Jamming via Energy Harvesting in UAV-Disrupted Channels}

\begin{document}
\author[1]{Ngoc-Tan Nguyen}
\author[2]{Thi-Thu Hoang}
\author[1]{Trung-Dung Hoang}
\author[1]{Thai-Duong Nguyen}

\affil[1]{\textit{VNU - University of Engineering and Technology, Hanoi, Vietnam}}
\affil[2]{\textit{Posts and Telecommunications Institute of Technology, Hanoi, Vietnam}}

\maketitle
\begin{abstract}
The open and broadcast nature of wireless communication systems, while enabling ubiquitous connectivity, also exposes them to jamming attacks that may critically compromise network performance or disrupt service availability. The proliferation of Unmanned Aerial Vehicles (UAVs) introduces a new dimension to this threat, as UAVs can act as mobile, intelligent jammers capable of launching sophisticated attacks by leveraging Line-of-Sight (LoS) channels and adaptive strategies. This paper addresses a critical challenge of countering intelligent UAV jamming in the context of energy-constrained ambient backscatter communication systems. Traditional anti-jamming techniques often fall short against such dynamic threats or are unsuitable for low-power backscatter devices. Hence, we propose a novel anti-jamming framework based on Deep Reinforcement Learning (DRL) that empowers the transmitter to not only defend against but also strategically exploit the UAV's jamming signals. In particular, our approach allows the transmitter to learn an optimal policy for switching between active transmission, energy harvesting from the jamming signal, and backscattering information using the jammer's own emissions. We then formulate the problem as a Markov Decision Process (MDP) and employ a Deep Q-Network (DQN) to derive the optimal operational strategy. Simulation results demonstrate that our DQN-based method significantly outperforms conventional Q-learning in convergence speed and surpasses a greedy anti-jamming strategy in terms of average throughput, packet loss rate, and packet delivery ratio.
\end{abstract}

\section{Introduction}
\label{intro}
Wireless communication technologies have become indispensable, underpinning a vast array of applications that permeate every facet of modern life. However, the broadcast nature of wireless channels makes them inherently vulnerable to malicious interference, particularly jamming attacks. A jammer can intentionally transmit disruptive signals to degrade the Signal-to-Interference-plus-Noise Ratio (SINR) at legitimate receivers, thereby interrupting or completely blocking valid communication links \cite{Pirayesh2022}, \cite{Vadlamani2016}. Unlike unintentional interference, jamming is often characterized by high power and persistence, posing a significant threat to network reliability and availability.

Recently, the emergence and widespread adoption of Unmanned Aerial Vehicles (UAVs) have presented both new opportunities and challenges for wireless networks. While UAVs can enhance network capacity and provide rapid deployment solutions \cite{Zeng2016}, especially in disaster-stricken or inaccessible areas, their agility and ability to establish strong Line-of-Sight (LoS) links also make them potent platforms for mobile jamming attacks \cite{Gao2020}. UAV-based jammers can dynamically adjust their positions and strategies to maximize disruption, posing a more severe threat than ground-based static jammers \cite{Li2019}. Addressing this evolving threat of intelligent UAV jamming is therefore crucial for ensuring the Quality of Service (QoS) and security of wireless communications.

Traditional anti-jamming techniques, such as power control, frequency hopping spread spectrum (FHSS) \cite{Proakis2000}, and rate adaptation, have been extensively studied. However, these methods often assume static or predictable jamming behaviors and may struggle against intelligent jammers that can learn and adapt their strategies \cite{Hanawal2014}. Moreover, for emerging low-power communication paradigms like ambient backscatter communication (AmBC) \cite{Liu2013}, energy efficiency is paramount, and conventional active anti-jamming responses might be too power-intensive. AmBC allows devices to communicate by reflecting existing ambient RF signals (e.g., TV, Wi-Fi, or even jamming signals), offering an ultra-low-power communication alternative. This is often coupled with Simultaneous Wireless Information and Power Transfer (SWIPT) principles, where devices can harvest energy from the same RF signals used for communication \cite{Jayakody2018}

While some prior works have explored reinforcement learning (RL) for anti-jamming in conventional wireless networks \cite{Hoang2023}, the problem of intelligent UAV jammers that strategically vary their power and presence, coupled with a system that can opportunistically harvest energy and backscatter using the jamming signal itself, presents unique challenges and opportunities. Many existing DRL-based anti-jamming solutions focus on channel switching or power adaptation in active transmission systems. For instance, Gao et al. \cite{Gao2020} proposed a DQN approach for anti-UAV jamming by optimizing transmission parameters. Van Huynh et al. \cite{VanHuynh2019} explored DRL for AmBC systems to defeat jammers, but the dynamic exploitation of varying UAV jamming signal strengths for adaptive energy harvesting and backscattering decisions needs further investigation. The strategic interplay between actively transmitting, harvesting energy for future use, or immediately backscattering data using the current jamming signal, all while facing an intelligent UAV jammer, forms the core of our investigation.

This paper proposes a novel anti-jamming framework for AmBC systems under intelligent UAV jamming attacks, leveraging Deep Reinforcement Learning (DRL) \cite{SuttonBarto2018}. Our key contributions are:
\begin{itemize}
    \item We design a system model where a transmitter, equipped with AmBC and energy harvesting capabilities, faces an intelligent UAV jammer. The transmitter can choose to: (i) actively transmit packets using stored energy when the channel is clear, (ii) harvest energy from the UAV's jamming signal, or (iii) backscatter its data using the ongoing jamming signal. This multifaceted response capability is crucial for adapting to the dynamic jamming environment.
    \item We formulate the anti-jamming decision-making process as a Markov Decision Process (MDP). The state space captures the jamming status, the transmitter's data buffer occupancy, and its stored energy level. The action space includes the operational modes mentioned above, as well as rate adaptation for active transmission.
    \item We develop a DRL-based solution using a Deep Q-Network (DQN) to learn the optimal policy that maximizes the long-term average throughput of the system. The DQN enables the transmitter to implicitly learn the UAV jammer's strategy and the environmental dynamics without explicit knowledge.
    \item We conduct extensive simulations to evaluate the performance of our proposed DQN-based approach. We show its superior convergence compared to traditional Q-learning and demonstrate its significant performance gains in terms of throughput, packet loss, and packet delivery ratio when benchmarked against a fixed greedy anti-jamming strategy under various jamming scenarios.
\end{itemize}
The remainder of this paper is organized as follows. Section \ref{sec:system_model} details the system model and formulates the problem. Section \ref{sec:solution} presents the proposed DRL-based anti-jamming solution. Section \ref{sec:evaluation} discusses the performance evaluation and simulation results. Finally, Section \ref{sec:conclusion} concludes the paper.

\section{System Model and Problem Formulation}
\label{sec:system_model}

\begin{figure}[t]
    \centering
    \includegraphics[width=1\linewidth]{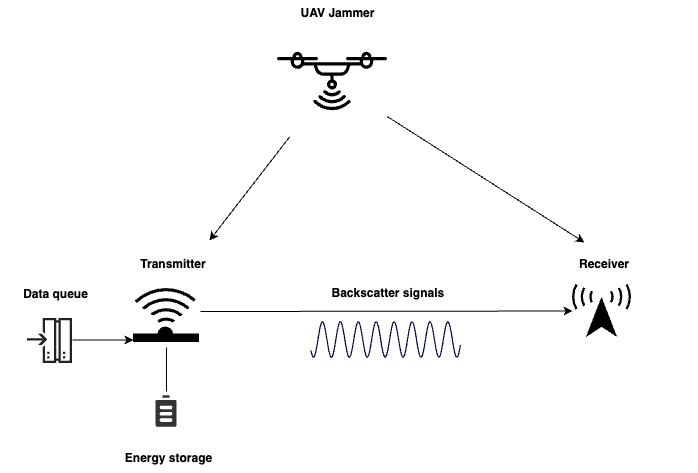}
    \caption{The system model, illustrating the interaction between the multi-modal transmitter, the receiver, and the intelligent UAV jammer.}
    \label{fig:sys_model}
\end{figure}

To effectively address the complex challenge of intelligent Unmanned Aerial Vehicle (UAV) jamming, we begin by establishing a comprehensive system model, which is conceptually depicted in Fig.~\ref{fig:sys_model}. This model serves as the analytical foundation for our subsequent problem formulation. 
\subsection{System Model}
We consider a communication system composed of three principal entities: a single transmitter (Tx), its intended receiver (Rx), and a malicious UAV jammer (J). At the core of this system, the transmitter is designed as a resource-constrained yet versatile device, specifically engineered to maintain functionality in hostile electromagnetic environments. This versatility is crucial for its survival and performance. The transmitter is equipped with a data buffer of a finite size, denoted as $D_{max}$, which is essential for managing the stochastic nature of packet arrivals, assumed to follow a Poisson process with an average rate of $\lambda$. Furthermore, it possesses an energy storage unit with a maximum capacity of $E_{max}$, representing its finite power reserve. Critically, the transmitter's resilience stems from two advanced operational modules: an energy harvesting (EH) module to scavenge power from ambient radio frequency (RF) signals, and an ambient backscatter communication (AmBC) circuit \cite{Liu2013}. The AmBC capability is particularly noteworthy as it allows the transmitter to communicate with ultra-low power consumption by modulating and reflecting incident RF signals (including the jammer's own disruptive signal)towards the receiver. This rich, multi-modal functionality enables a diverse set of strategic responses. When confronted with the dynamic threat of jamming, the transmitter can dynamically select its operational mode from several distinct choices: performing a conventional Active Transmission (AT) by consuming its stored energy; entering an EH mode to replenish its energy reserves by capitalizing on the jammer's signal; leveraging AmBC for power-efficient data transfer; applying Rate Adaptation (RA) to enhance the robustness of its active transmissions; or remaining Idle to conserve its limited resources. The ability to intelligently switch between these modes forms the central premise of our proposed defense mechanism.

\subsection{Channel Model}
Having defined the capabilities of our defensive transmitter, we now turn to characterizing the adversary it is designed to overcome: the intelligent UAV jammer. The jammer's principal objective is to disrupt the Tx-Rx communication link by minimizing the Signal-to-Interference-plus-Noise Ratio (SINR) at the receiver. The SINR, a fundamental metric for communication quality denoted by $\theta$, is expressed as:
\begin{equation}
    \theta = \frac{P_R}{P_{J,\text{eff}} + \sigma^2},
    \label{eq:sinr}
\end{equation}
where $P_R$ represents the received signal power from the transmitter, $P_{J,\text{eff}}$ is the effective jamming power impacting the receiver, and $\sigma^2$ denotes the power of Additive White Gaussian Noise (AWGN). The jammer's intelligence is manifested in its ability to dynamically adjust its physical location and transmission power, $P_J$, to maximize its disruptive impact. Its high mobility and potential to establish strong Line-of-Sight (LoS) links make it a particularly potent threat, causing the effective jamming power to fluctuate significantly and creating a highly non-stationary channel environment. The air-to-ground (ATG) channel between the UAV and the transmitter is subject to both large-scale and small-scale fading. Al-Hourani et al describe the path loss (PL), based on the distance $d$ between the UAV and the transmitter, can be modeled as \cite{AlHourani2014}:
\begin{equation}
    \text{PL}(d) = 
    \begin{cases} 
        \beta_{\text{LoS}} d^{-\alpha} & \text{for Line-of-Sight (LoS)} \\
        \beta_{\text{NLoS}} d^{-\alpha} & \text{for Non-Line-of-Sight (NLoS)}
    \end{cases}
    \label{eq:pathloss}
\end{equation}
where $\alpha$ is the path loss exponent, and $\beta_{\text{LoS}}$, $\beta_{\text{NLoS}}$ are additional attenuation factors for LoS and NLoS links, respectively. The probability of having an LoS connection, $P_{\text{LoS}}$, depends on the environment and the elevation angle $\theta_i$ (in degrees) between the nodes given by \cite{AlHourani2014}:
\begin{equation}
    P_{\text{LoS}}(\theta_i) = \frac{1}{1 + \Phi \exp(-\Psi[\theta_i - \Phi])},
\end{equation}
where $\Phi$ and $\Psi$ are environment-dependent constants. 
To capture this, we model the fluctuating power as taking discrete levels from a set $\mathcal{P}_J = \{P_0^J, P_1^J, \ldots, P_N^J\}$, where $P_0^J = 0$W signifies the absence of effective jamming. A crucial aspect of our model, reflecting real-world conditions, is the information asymmetry between the transmitter and the jammer. The transmitter operates under partial observability; it can reliably detect whether jamming is present ($j=1$) or absent ($j=0$), but it cannot ascertain the specific power level of the jamming signal. This forces the transmitter to make decisions under significant uncertainty, rendering traditional deterministic anti-jamming policies ineffective and necessitating a robust, learning-based approach. The interaction between the transmitter and the jammer is thus governed by the perceived state of the communication channel. When the channel is clear ($j=0$), the transmitter has the opportunity to actively transmit up to $\hat{d}_t$ packets, provided it has sufficient data in its buffer and energy in its storage unit. Conversely, during a jamming state ($j=1$), the transmitter faces a profound strategic dilemma. It must intelligently choose between attempting a robust active transmission via rate adaptation, entering EH mode to convert the hostile jamming signal into useful energy, or using the AmBC mode to transmit data passively. The efficacy of both the EH and AmBC actions is directly contingent upon the unknown jamming power level $P_n^J \in \mathcal{P}_J$, further complicating the decision-making process.

\subsection{MDP Problem Formulation}
To develop an optimal defense strategy that can navigate these complex and uncertain conditions, we formally structure the transmitter's sequential decision-making problem as a Markov Decision Process (MDP). The MDP provides a standard and powerful mathematical framework for modeling goal-directed learning through interaction \cite{SuttonBarto2018}, and is defined by a tuple containing the state space, action space, reward function, and the overarching objective. A precise definition of these components is essential for applying a learning algorithm. The system state at any given time slot $t$, denoted as $s_t$, must encapsulate all necessary information for making an optimal decision. Accordingly, the state representation $s_t = (j_t, d_t, e_t)$ is carefully chosen to be both compact and sufficiently informative. Here, $j_t \in \{0, 1\}$ represents the binary jamming status which dictates the available actions, $d_t \in \{0, 1, \ldots, D_{max}\}$ indicates the current number of packets in the data buffer which reflects the urgency to communicate, and $e_t \in \{0, 1, \ldots, E_{max}\}$ specifies the current energy level, which represents the transmitter's capacity for future actions. 
Formally, the state space $\mathcal{S}$ is:
\begin{equation}
  \mathcal{S}
  = \{\, (j,d,e) \mid
        j\!\in\!\{0,1\},
        \, d\!\in\![0,D_{\max}],
        \, e\!\in\![0,E_{\max}] \,\}.
  \label{eq:state_space}
\end{equation}

In response to observing the state $s_t$, the transmitter must select an action $a_t$ from a rich action space $\mathcal{A}$ that reflects its versatile hardware capabilities: Idle, AT, EH, AmBC, and $M$ distinct levels of AT-RA. Each action represents a strategic trade-off; for instance, AT offers high throughput but is energy-intensive, AmBC is energy-efficient but opportunistic, and EH is an investment in future transmission capability. The action space is then defined as:
\begin{equation}
\label{eq:action_space_split}
a_t \in \mathcal{A} = \bigl\{
  \begin{aligned}[t]
    &\underbrace{\text{Idle}}_{1},\;
     \underbrace{\text{AT}}_{2},\;
     \underbrace{\text{EH}}_{3},\;
     \underbrace{\text{AmBC}}_{4},\\
    &\underbrace{\text{AT-RA}_1,\ldots,\text{AT-RA}_M}_{4{+}1,\ldots,4{+}M}
  \end{aligned}
\bigr\}.
\end{equation}

The immediate consequence of an action is quantified by a reward function, $r(s_t, a_t)$, defined as the number of successfully delivered packets. For instance, the reward for taking action $a$ in state $s=(j,d,e)$ can be expressed as:
\begin{equation}
    r(s,a) = 
    \begin{cases} 
        d_t & \text{if } a=\text{AT} \text{ and } j=0, e \ge e_t \\
        d_{n}^J & \text{if } a=\text{AmBC} \text{ and } j=1 \\
        d_{m}^r & \text{if } a=\text{AT-RA}_m \text{ and } j=1, e \ge e_t \\
        0 & \text{otherwise},
    \end{cases}
    \label{eq:reward_func}
\end{equation}
where $d_t$, $d_{n}^J$, and $d_{m}^r$ are the packets successfully sent in each respective case. The ultimate goal is to find an optimal policy, $\pi^*: \mathcal{S} \rightarrow \mathcal{A}$, that maximizes the cumulative long-term average system throughput, denoted as $R(\pi)$. This objective, which captures the desire for sustained performance and resilience, is formally expressed as:
\begin{equation}
    \max_{\pi} R(\pi) = \lim_{T\rightarrow\infty} \frac{1}{T} \sum_{k=1}^{T} \mathbb{E}[r_k(s_k, \pi(s_k))].
    \label{eq:objective}
\end{equation}
Given that the state space is finite and the underlying transition probabilities form an irreducible Markov chain, this long-term average reward is well-defined and independent of the system's starting state, ensuring a consistent and meaningful optimization goal.

\section{Proposed DRL-based Anti-Jamming Solution}
\label{sec:solution}
Having formulated the problem as an MDP, this section details the intelligent learning-based solution we propose to solve it. To properly evaluate its effectiveness, we first introduce a traditional heuristic approach that will serve as a performance benchmark.
\subsection{Benchmark: A Heuristic Greedy Strategy}
As a point of comparison for our learning-based approach, we first define a heuristic, rule-based greedy strategy. This strategy is designed to be intuitive and simple to implement, reflecting a common approach to such problems without the overhead of machine learning. The policy is bifurcated based on the perceived channel condition. Firstly, in the absence of jamming ($j=0$), the transmitter adopts a straightforward aggressive policy: it will always attempt to perform Active Transmission to send as many packets as its energy and buffer state allow. Secondly, when jamming is detected ($j=1$), the transmitter follows a fixed, alternating cycle of opportunistic actions. It will perform Energy Harvesting for a predefined period to gather power from the hostile signal, and then switch to Ambient Backscattering for the remainder of the cycle to transmit data. 

While simple, this greedy strategy suffers from significant drawbacks that limit its effectiveness in a dynamic environment. Its policy is static and rule-based, rendering it incapable of adapting to the changing tactics of an intelligent jammer. For instance, its performance is highly sensitive to manually-tuned hyperparameters, such as the harvesting period ($T_{harvest}$), which may be optimal for one jamming pattern but highly suboptimal for another. Furthermore, its deterministic nature makes it predictable and thus, easily exploitable by a sophisticated adversary that can learn its response pattern. These inherent limitations motivate our proposal for a more intelligent and adaptive solution.

\subsection{Intelligent Solution: Deep Q-Network}
To overcome the inherent rigidity and sub-optimality of the greedy baseline, we propose an intelligent solution based on Deep Reinforcement Learning (DRL), a paradigm that has demonstrated remarkable success in solving complex decision-making problems under uncertainty \cite{Hoang2023}. DRL is exceptionally well suited to this problem because it allows the agent, in this case the transmitter, to learn effective strategies through direct trial-and-error interaction with its environment, thereby eliminating the need for an explicit environmental model. Specifically, we employ the Deep Q-Network (DQN) algorithm, a seminal value-based DRL method renowned for its ability to solve problems with high-dimensional state spaces where traditional tabular methods like Q-learning would fail due to the curse of dimensionality \cite{Watkins1992}. Instead of maintaining an exhaustive and memory-intensive lookup table for state-action values, the DQN algorithm utilizes a deep neural network, parameterized by a set of weights $\theta$, as a powerful non-linear function approximator to estimate the action-value function, $Q(s, a; \theta)$.

To ensure that the learning process, which involves training a deep neural network with potentially correlated data, is both stable and efficient, the DQN framework integrates two architectural cornerstones: experience replay and a target network. The first technique, experience replay, addresses the issue of sample correlation by storing the agent’s experiences, represented as transitions $(s_t, a_t, r_t, s_{t+1})$, in a large but finite replay-memory buffer denoted $\mathcal{D}$. During the training phase, the algorithm does not learn from experiences as they occur sequentially; instead, it samples random mini-batches of transitions from this buffer. This practice effectively breaks the temporal correlations in the data stream, which is crucial for the stability of the gradient-based optimization used to train the neural network \cite{Lin1992}. The next technique, which complements experience replay, is the use of a separate target network to address the "moving target" problem. A target network, $\hat{Q}(s, a; \theta')$, which is a periodic clone of the primary Q-network $Q(s, a; \theta)$, is used to generate the target values for the learning update. Its weights, $\theta'$, are kept frozen for a fixed number of iterations and are only then updated with the weights from the primary network. This provides a consistent and stable target, $y_i$, for the Bellman update, preventing the oscillatory or divergent behavior that can occur when a network tries to chase its own rapidly changing predictions. This technique, combined with innovations like Double Q-learning which mitigates overestimation bias \cite{VanHasselt2016}, enhances learning stability. Further improvements, such as Dueling Network Architectures \cite{Wang2016}, could also be integrated to separate the estimation of state values and action advantages. The target value for a given transition $i$ from the mini-batch is calculated as:
\begin{equation}
    y_i = r_i + \gamma \max_{a'} \hat{Q}(s'_{i}, a'; \theta'),
\end{equation}
where $\gamma$ is the discount factor that balances the importance of immediate versus future rewards. The primary network is then trained by minimizing the Mean Squared Error (MSE) loss between its predicted Q-values and these stable target values, using an efficient optimizer like Adam \cite{Kingma2014}:
\begin{equation}
    L(\theta_i) = \mathbb{E}_{(s,a,r,s') \sim U(\mathcal{D})} \left[ \left( y_i - Q(s_i, a_i; \theta_i) \right)^2 \right].
\end{equation}
A final crucial element of the learning process is managing the exploration-exploitation trade-off \cite{SuttonBarto2018}. This dilemma is fundamental to reinforcement learning: the agent must exploit actions that it has learned are effective, but it must also explore new, seemingly suboptimal actions to discover potentially even better long-term strategies. To navigate this, we employ an $\epsilon$-greedy policy. With a small probability $\epsilon$, the agent selects a random action, thereby exploring the action space. Otherwise, with probability $1-\epsilon$, it selects the action with the highest currently estimated Q-value, thereby exploiting its knowledge. The value of $\epsilon$ is typically annealed over the course of training, starting high to encourage broad exploration and gradually decreasing to favor exploitation as the agent becomes more confident in its learned policy. The synergy of these components—the neural network approximator, experience replay, the target network, and the $\epsilon$-greedy exploration strategy—is encapsulated in the comprehensive DQN-based anti-jamming process outlined in Algorithm~\ref{alg:dqn_anti_jamming}.

\begin{algorithm}[H]
\caption{DQN-based Anti-Jamming Algorithm}
\label{alg:dqn_anti_jamming}
\begin{algorithmic}[1]
\State Initialize replay memory $\mathcal{D}$, Q-network $Q$ with weights $\theta$, target network $\hat{Q}$ with weights $\theta' \leftarrow \theta$.
\For{episode = 1 to $T_{episodes}$}
    \State Initialize state $s_1$.
    \For{t = 1 to $T_{steps}$}
        \State Select action $a_t$ using $\epsilon$-greedy policy based on $Q(s_t, a; \theta)$.
        \State Execute $a_t$, observe reward $r_t$ and next state $s_{t+1}$.
        \State Store transition $(s_t, a_t, r_t, s_{t+1})$ in $\mathcal{D}$.
        \If{replay memory is sufficiently full}
            \State Sample a random mini-batch from $\mathcal{D}$.
            \State Compute target values $y_j$ for the mini-batch.
            \State Perform a gradient descent step on the loss $L(\theta)$.
        \EndIf
        \State Periodically update target network: $\theta' \leftarrow \theta$.
    \EndFor
\EndFor
\end{algorithmic}
\end{algorithm}

\section{Performance Evaluation}
\label{sec:evaluation}
In this section, we evaluate the performance of our proposed DQN-based anti-jamming scheme. We first describe the simulation setup and parameters, then compare the convergence of DQN with traditional Q-learning, and finally benchmark DQN against a greedy anti-jamming strategy.

\subsection{Simulation Setup}
The system parameters are set as follows:
\begin{table}[H]
\centering
\caption{Simulation System Parameters}
\label{tab:sim_params}
\begin{tabular}{@{}ll@{}}
\toprule
\textbf{Parameter} & \textbf{Value} \\ \midrule
\multicolumn{2}{l}{\textit{\textbf{System Parameters}}} \\
Max Buffer Size ($D_{max}$) & 10 packets \\
Max Energy Storage ($E_{max}$) & 10 units \\
Packet Arrival Rate ($\lambda$) & 3 packets/slot (Poisson) \\
Active Tx Packets (No Jamming, $\hat{d}_t$) & 4 packets \\
Active Tx Energy Cost ($e_t$) & 1 unit/packet \\
Jamming Power Levels ($\mathcal{P}_J$) & \{0, 5, 10, 15\} W \\
Default Average Jamming Power ($P_{avg}$) & 7 W \\
Harvested Energy ($e^J$) & \{0, 1, 2, 3\} units \\
Backscatterable Packets ($\hat{d}^{AmBC}$) & \{0, 1, 2, 3\} packets \\
\bottomrule
\end{tabular}
\end{table}

\begin{table}[H]
\centering
\caption{Simulation DQN Parameters}
\label{tab:dqn_params}
\begin{tabular}{@{}ll@{}}
\toprule
\multicolumn{2}{l}{\textit{\textbf{DQN Parameters}}} \\
Learning Rate ($\alpha$) & 0.0001 \\
Discount Factor ($\gamma$) & 0.9 \\
Replay Memory Size & 10000 \\
Batch Size & 32 \\
Target Network Update Freq. (C) & 5000 steps \\
$\epsilon$-Greedy Schedule & Start 1.0, End 0.01, Decay 0.9999 \\
\bottomrule
\end{tabular}
\end{table}

\subsection{Convergence Comparison: DQN vs. Q-learning}
\begin{figure}[H]
    \centering
    \includegraphics[width=1\linewidth]{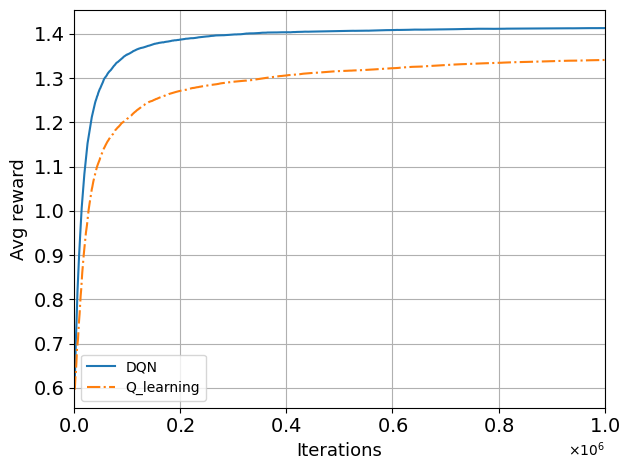}
    \caption{Convergence comparison of DQN and Q-learning.}
    \label{fig:convergence}
\end{figure}

Fig.~\ref{fig:convergence} compares the convergence of the DQN agent against a standard tabular Q-learning agent. The average reward per episode is plotted against the number of training iterations. The DQN agent achieves a higher and more stable reward much faster than Q-learning. This is attributed to DQN's ability to generalize across the large state space using its neural network, whereas Q-learning must visit each state-action pair sufficiently often, which is inefficient.

\subsection{Performance Comparison with Greedy Strategy}

\subsubsection{Impact of Average Jamming Power ($P_{avg}$)}
\begin{figure}[H]
    \centering
    \includegraphics[width=1\linewidth]{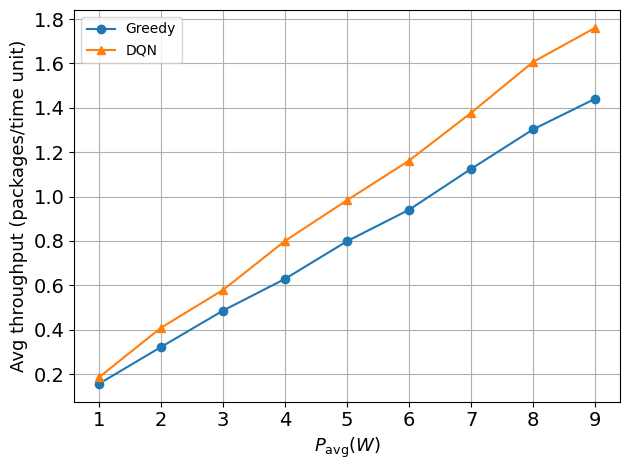}
    \caption{Average throughput (DQN vs. Greedy) vs. $P_{avg}$}
    \label{fig:perf_vs_p_avg_throughput}
\end{figure}

\begin{figure}[H]
    \centering
    \includegraphics[width=1\linewidth]{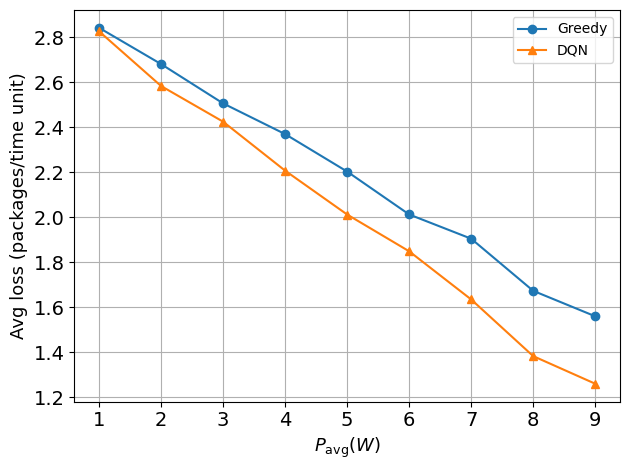}
    \caption{Average packet loss (DQN vs. Greedy) vs. $P_{avg}$}
    \label{fig:perf_vs_p_avg_loss}
\end{figure}

\begin{figure}[H]
    \centering
    \includegraphics[width=1\linewidth]{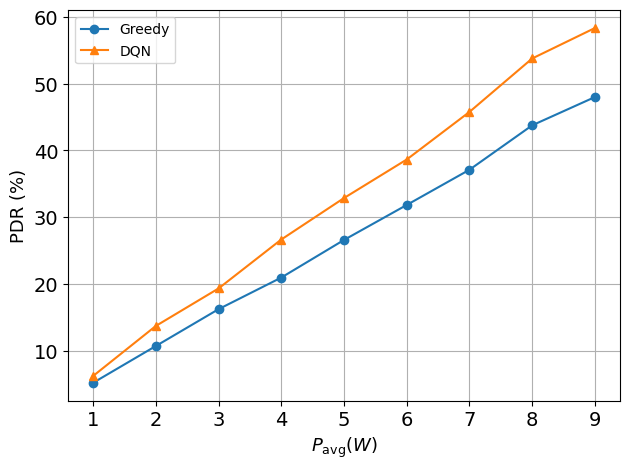}
    \caption{PDR (DQN vs. Greedy) vs. $P_{avg}$}
    \label{fig:perf_vs_p_avg_pdr}
\end{figure}

Figs.~\ref{fig:perf_vs_p_avg_throughput}, \ref{fig:perf_vs_p_avg_loss}, and \ref{fig:perf_vs_p_avg_pdr} show the system performance as the average jamming power of the UAV varies. As $P_{avg}$ increases, the DQN agent achieves significantly higher throughput and PDR, and lower packet loss, compared to the greedy baseline. This highlights DQN's core advantage: it learns to exploit the jamming signal. With stronger jamming, the potential reward from both EH (for future AT) and AmBC increases. The DQN agent learns the optimal trade-off between these choices, effectively turning the stronger threat into a more valuable resource. The greedy strategy, locked into a fixed EH/AmBC cycle, cannot adapt and thus fails to capitalize as effectively.



\subsubsection{Impact of Active Transmission Capability ($\hat{d}_t$)}
\begin{figure}[H]
    \centering
    \includegraphics[width=1\linewidth]{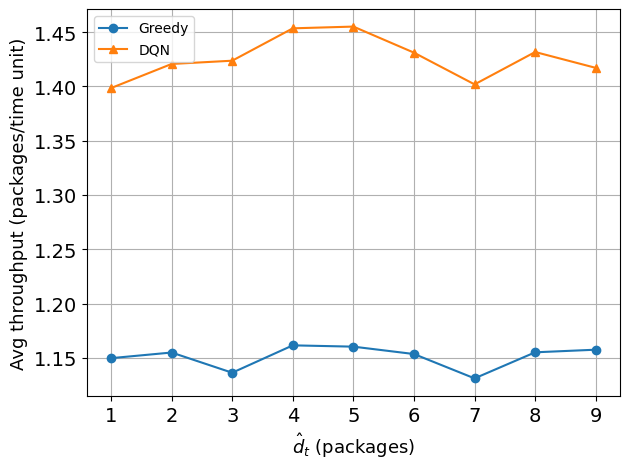}
    \caption{Average throughput (DQN vs. Greedy) vs. $\hat{d}_t$}
    \label{fig:perf_vs_p_avg_throughput_dt}
\end{figure}

\begin{figure}[H]
    \centering
    \includegraphics[width=1\linewidth]{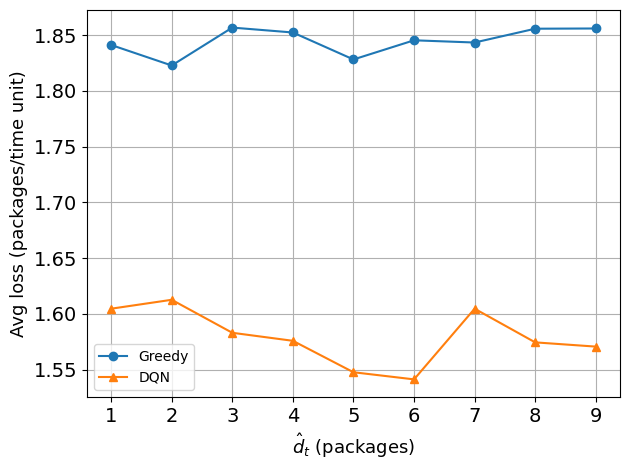}
    \caption{Average packet loss (DQN vs. Greedy) vs. $\hat{d}_t$}
    \label{fig:perf_vs_p_avg_loss_dt}
\end{figure}

\begin{figure}[H]
    \centering
    \includegraphics[width=1\linewidth]{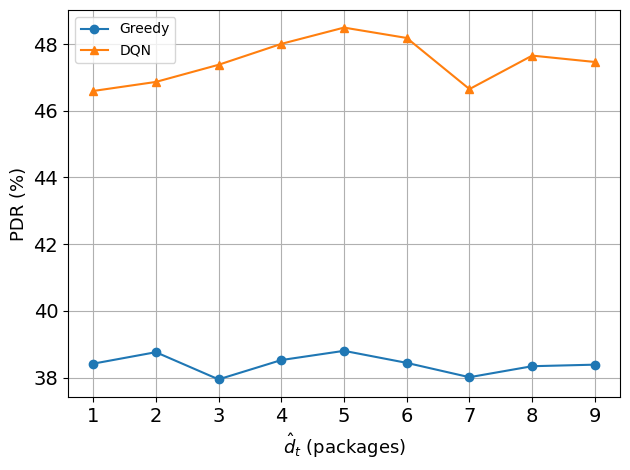}
    \caption{PDR (DQN vs. Greedy) vs. $\hat{d}_t$}
    \label{fig:perf_vs_p_avg_pdr_dt}
\end{figure}

Figs.~\ref{fig:perf_vs_p_avg_throughput_dt}, \ref{fig:perf_vs_p_avg_loss_dt}, and \ref{fig:perf_vs_p_avg_pdr_dt} illustrate performance for varying $\hat{d}_t$ (the max packets for AT in a clear channel) at a fixed $P_{avg}=7$W. While a higher $\hat{d}_t$ benefits both strategies by improving performance during non-jamming periods, the DQN agent consistently maintains a significant performance gap. This shows that the DQN's learned policy is robust and effective regardless of the baseline active transmission capacity, as its strength lies in optimizing behavior during the more complex jamming periods.
    
\subsubsection{Impact of Greedy Strategy's Harvesting Period ($T_{harvest}$)}
\begin{figure}[H]
    \centering
    \includegraphics[width=1\linewidth]{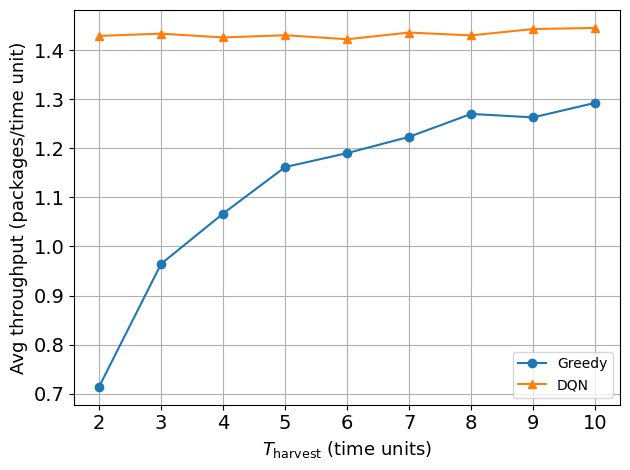}
    \caption{Average throughput (DQN vs. Greedy) vs. $T_{harvest}$}
    \label{fig:perf_vs_p_avg_throughput_T_harvest}
\end{figure}

\begin{figure}[H]
    \centering
    \includegraphics[width=1\linewidth]{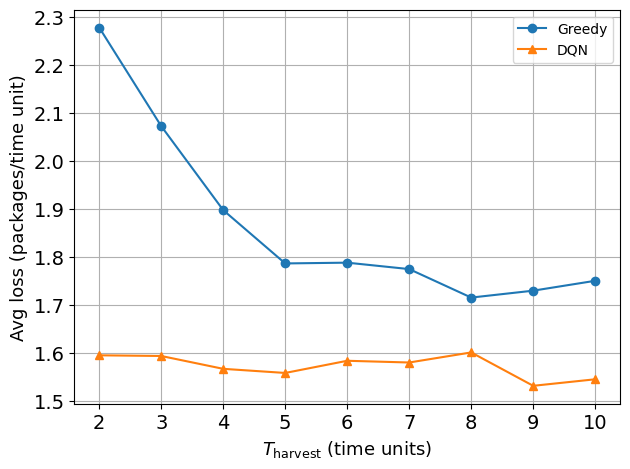}
    \caption{Average packet loss (DQN vs. Greedy) vs. $T_{harvest}$}
    \label{fig:perf_vs_p_avg_loss_T_harvest}
\end{figure}

\begin{figure}[H]
    \centering
    \includegraphics[width=1\linewidth]{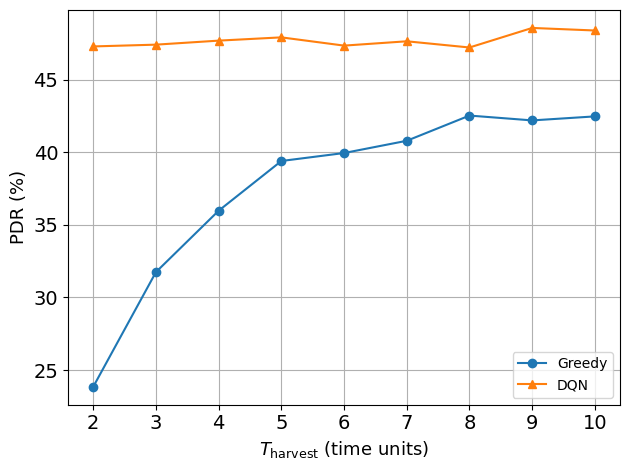}
    \caption{PDR (DQN vs. Greedy) vs. $T_{harvest}$}
    \label{fig:perf_vs_p_avg_pdr_T_harvest}
\end{figure}

Figs.~\ref{fig:perf_vs_p_avg_throughput_T_harvest}, \ref{fig:perf_vs_p_avg_loss_T_harvest}, and \ref{fig:perf_vs_p_avg_pdr_T_harvest} show the impact of varying the fixed harvesting period $T_{harvest}$ for the greedy strategy, again at $P_{avg}=7$W. The greedy strategy's performance is highly sensitive to this hyperparameter, achieving optimal performance at a specific value before declining. The DQN agent, however, outperforms the greedy strategy even at its optimal setting. This result powerfully illustrates the advantage of a learning-based adaptive policy over a fixed, rule-based heuristic, which requires manual tuning and cannot adapt to changing environmental statistics.

\section{Conclusion}
\label{sec:conclusion}

This paper has addressed the critical challenge of intelligent UAV jamming in ambient backscatter communication systems by proposing a novel anti-jamming strategy based on Deep Q-Networks. Through the formulation of the problem as a MDP, we have enabled the transmitter to learn a sophisticated policy that adaptively switches between active transmission, energy harvesting, and ambient backscattering. The extensive simulation results have confirmed that the proposed DQN approach significantly outperforms a static greedy baseline in key performance metrics. The principal finding of this work is the demonstrated ability of the learning agent to transform an adversarial jamming signal into an opportunistic resource, thereby enhancing system resilience and throughput. Future work can extend this framework to explore multi-agent DRL scenarios and more complex jamming strategies, potentially incorporating physical layer security metrics.

\section{Acknowledgment}
This work has been supported by Posts and Telecommunications Institute of Technology under the project number 07-2025-HV-VT1.



\end{document}